\documentclass{article}
\usepackage{spconf,amsmath,graphicx,hyperref}

\usepackage{longtable}
\usepackage{amssymb, booktabs}
\usepackage{anyfontsize}
\usepackage{multirow}
\usepackage{caption}
\usepackage[table]{xcolor}
\usepackage{float}
\usepackage{url}
\usepackage{cite}
\usepackage{threeparttable}
\DeclareCaptionFont{mycaptionsize}{\fontsize{9pt}{10pt}\selectfont}
\title{MG-REPA: Multi-Granularity Representation Alignment for Generative Target Speaker Extraction}
\name{Dohwan Kim$^{1}$ \qquad Jung-Woo Choi$^{2 *}$\thanks{$^*$Corresponding author.}}
\address{$^{1}$School of Electrical Engineering, KAIST, Daejeon, South Korea \\
$^{2}$Department of AI Systems, KAIST, Daejeon, South Korea \\
\{dohwan, jwoo\}@kaist.ac.kr}

\begin{document}
\ninept
\maketitle

\begin{abstract}
\noindent Generative target speaker extraction (TSE) models achieve high-quality extraction, but suffer from considerably slow training convergence. Representation Alignment (REPA) accelerates diffusion transformer training by aligning hidden states with a pre-trained Self-Supervised Learning (SSL) teacher. However, existing alignment in the speech and audio domain operates at a single granularity. For TSE, frame-level alignment leaves the global, time-invariant characteristics of the target speech implicit, whereas utterance-level alignment leaves the time-varying content implicit, even though extracting the target requires both. We propose Multi-Granularity Representation Alignment (MG-REPA), aligning the hidden states with an SSL model at the frame level and a speaker verification model at the utterance level. We further apply temporal normalization to the frame-level target, removing its utterance-level statistics so that each branch concentrates on its granularity. On Libri2Mix, MG-REPA reaches the OVRL and PESQ of a MeanFlow-TSE baseline with $3.0\times$ and $2.1\times$ fewer epochs, respectively. Trained for 750 epochs, it is competitive with the published 2,000-epoch baseline on intrusive metrics while improving non-intrusive metrics. An ablation demonstrates that MG-REPA outperforms alignment at a single granularity.
 \footnote{\url{https://github.com/rlaehghks5/MG-REPA}}

\end{abstract}
\noindent
\vspace{-0.1cm}
\section{Introduction}
\label{sec:intro}
\vspace{-0.1cm}

In real-world multi-talker environments, isolating a target voice from mixtures of overlapping speech and background noise is a fundamental challenge. Target Speaker Extraction (TSE) recovers the target speech using auxiliary clues such as a short reference enrollment utterance~\cite{VoiceFilter, Spex+}. Discriminative TSE models achieve strong signal fidelity but often introduce unnatural artifacts and generalize poorly~\cite{NeuralTSE, DDTSE}. To mitigate these limitations, generative formulations built on diffusion models~\cite{ho2020denoising} and flow matching~\cite{lipman2023flow} have been introduced. 
Diffusion-based TSE~\cite{DiffTSE2, DDTSE, SOLOSPEECH} attains high perceptual quality, but requires multiple sampling steps. Flow-based TSE learns straighter probability paths, and recent systems~\cite{flowtse, ADFlowTSE, MeanFlowTSE} have pushed inference down to a few or even a single step while retaining perceptual quality. Despite this fast inference, these one-step models still suffer from slow training convergence, requiring up to 2,000 epochs.


\begin{figure}
\begin{minipage}[b]{.48\linewidth}
  \centering
  \centerline{\includegraphics[width=4cm]{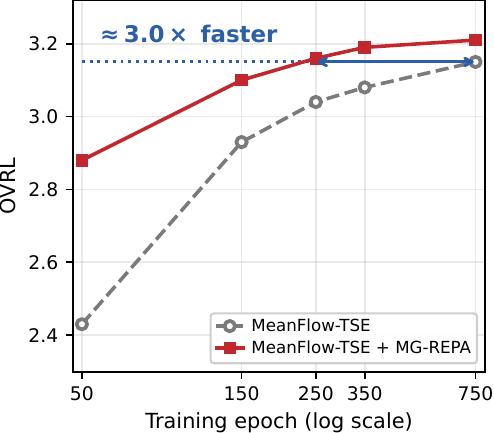}}
\end{minipage}
\begin{minipage}[b]{.48\linewidth}
  \centering
  \centerline{\includegraphics[width=4cm]{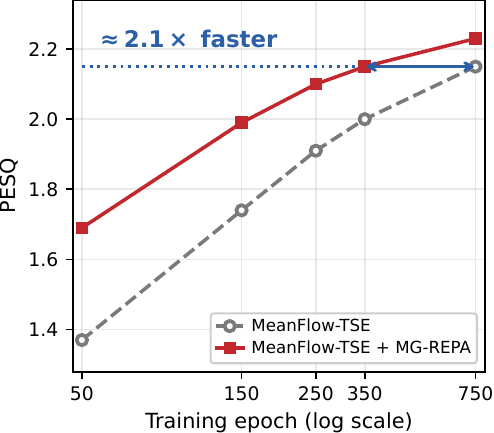}}
\end{minipage}
\vspace{0.2cm}
\caption{\textbf{MG-REPA significantly accelerates diffusion transformer training.} Convergence of OVRL and PESQ on Libri2Mix Noisy.}
\label{fig:convergence}
\vspace{-0.6cm}
\end{figure}

To accelerate the training of generative models, Representation Alignment (REPA)~\cite{REPA} was introduced. Identifying representation learning as a major bottleneck in diffusion transformer training, REPA adds a simple auxiliary regularization that aligns the projected hidden states of an intermediate layer with the representations extracted from the clean image by a frozen Self-Supervised Learning (SSL) encoder, yielding considerably faster convergence.

REPA has gained attention in the speech and audio domain~\cite{AG-REPA, AnyEnhance, A-DMA, ImmersiveTTS}, where the generator's intermediate features are aligned with representations from pre-trained speech or audio encoders~\cite{HUBERT, WAVLM, ATST}. Some of these works further show that combining multiple alignment objectives is beneficial. A-DMA~\cite{A-DMA} pairs speech representation alignment with a speech SSL teacher and a text-guided auxiliary CTC loss on the intermediate states, doubling the convergence speed of diffusion-based TTS. ImmersiveTTS~\cite{ImmersiveTTS} aligns the generator with both the speech-domain SSL model WavLM~\cite{WAVLM} and the audio-domain SSL model ATST-Frame~\cite{ATST}. AG-REPA~\cite{AG-REPA} instead aligns the generator at the utterance level, pooling the hidden states over time. 

However, each of these approaches aligns at a single granularity, either frame-level or utterance-level, even when multiple teachers are used. We argue that a single granularity is insufficient for TSE. A speech signal encompasses two complementary dimensions: (1) fine-grained, time-varying frame-level content, and (2) global, time-invariant utterance-level characteristics. Frame-level alignment explicitly supervises only the former and utterance-level alignment only the latter, even though both are important. 

To address this, we propose Multi-Granularity Representation Alignment (MG-REPA), which tailors the alignment to TSE by supervising the generator at both the frame and utterance levels. MG-REPA decouples the alignment into two branches attached to an intermediate layer of the generator: a frame-level REPA guided by a speech SSL model for fine-grained phonetic content, and an utterance-level REPA guided by a speaker verification model, whose utterance embedding summarizes these global characteristics. The two branches take the same hidden states as input but project them through separate adapters, so that each granularity is served by its own mapping. 

\noindent To remove utterance-level statistics from the frame-level targets, we further apply temporal normalization so that each branch focuses on a distinct granularity. We demonstrate that MG-REPA exceeds alignment at a single granularity. Moreover, on Libri2Mix~\cite{LIBRIMIX}, MG-REPA attains the 750-epoch OVRL and PESQ of a MeanFlow-TSE baseline retrained under the same recipe with 250 and 350 epochs, corresponding to $3.0\times$ and $2.1\times$ faster convergence, respectively. With 750 epochs, MG-REPA remains comparable to the published 2,000-epoch baseline on intrusive metrics and exceeds it on non-intrusive metrics.

\vspace{-0.2cm}
\section{Background}
\label{sec:background}
\vspace{-0.2cm}

\subsection{Flow-based TSE} \label{ssec:flow_tse} 
\vspace{-0.1cm}
Let $\mathbf{y} = \mathbf{s} + \mathbf{b} \in \mathbb{R}^{L}$ be a single-channel mixture of the target speech $\mathbf{s}$ and an interference noise $\mathbf{b}$. 
Given an enrollment utterance $\mathbf{e} \in \mathbb{R}^{L_{e}}$ identifying the target, the goal of Target Speaker Extraction (TSE) is to estimate $\hat{\mathbf{s}}$ from $\mathbf{y}$ and $\mathbf{e}$. We work in the complex STFT domain and denote $\mathbf{Y}, \mathbf{S} \in \mathbb{R}^{2F \times T}$ and $\mathbf{E} \in \mathbb{R}^{2F \times T_{e}}$ as the spectra of $\mathbf{y}$, $\mathbf{s}$, and $\mathbf{e}$, with $F$ frequency bins and $T$ and $T_e$ time frames for the mixture and the enrollment, respectively.

Flow matching~\cite{lipman2023flow} learns an instantaneous velocity field that transports a prior distribution $p_0$ to the data distribution $p_1$. A probability path is prescribed between a prior sample $\mathbf{z_0} \sim p_0$ and a data sample $\mathbf{z_1} \sim p_1$. Optimal Transport defines the probability path as a simple linear interpolation over time $t$:
\begin{equation} \mathbf{z}_{t} = (1-t)\, \mathbf{z_0} + t\, \mathbf{z_1}, \qquad t \in [0,1] \label{eq:ot_path} \end{equation} 

\noindent The generative process tracking this path is governed by
an Ordinary Differential Equation for the instantaneous velocity field: 
\begin{equation} \mathbf{v}_t(\mathbf{z}_{t}, t) = \frac{d\mathbf{z}_{t}}{dt} = \mathbf{z_1} - \mathbf{z_0} \label{eq:target_vel} \end{equation} 

\noindent A network $\mathbf{v}_{\theta}$, conditioned on the enrollment $\mathbf{E}$, is then trained to minimize the flow matching loss: 
\vspace{-0.1cm}

\begin{equation} \mathcal{L}_{\mathrm{FM}} = \mathbb{E}_{t, \mathbf{z}_{0}, \mathbf{z}_{1}} \!\left[ \big\| \mathbf{v}_{\theta}(\mathbf{z}_{t}, t, \mathbf{E}) - \mathbf{v}_t(\mathbf{z}_{t}, t) \big\|^{2} \right] \label{eq:fm} \end{equation}

\noindent The endpoints $\mathbf{z}_{0}$ and $\mathbf{z}_{1}$ are a design choice that determines where inference can start and end. FlowTSE~\cite{flowtse} sets $\mathbf{z}_{0}$ to Gaussian noise and $\mathbf{z}_{1}$ to the target speech, so generation starts from noise and spans the full interval. AD-FlowTSE~\cite{ADFlowTSE} instead places both endpoints on signal-domain quantities, $\mathbf{z}_{0} = \mathbf{B}$ and $\mathbf{z}_{1} = \mathbf{S}$, and models the mixture as a convex combination $\mathbf{y} = \lambda \mathbf{s} + (1-\lambda)\mathbf{b}$ with a mixing ratio $\lambda \in [0,1]$, so that the mixture already lies on the path, $\mathbf{Y} = \mathbf{z}_{\lambda}$. Inference can then be initialized at the mixture and integrated only over $[\hat{\lambda}, 1]$, where $\hat{\lambda}$ is supplied by a mixing-ratio predictor.

\vspace{-0.2cm}
\subsection{MeanFlow-based TSE} 
\vspace{-0.1cm}
\label{ssec:meanflow_tse}

Since the instantaneous velocity only determines the tangent direction of a generally curved trajectory, a coarse discretization along it yields inaccurate results, which MeanFlow~\cite{geng2025meanflow} avoids by modeling the average velocity over an interval $[t, r]$, where $0 \le t < r \le 1$, so that a single evaluation moves the state exactly from $\mathbf{z}_{t}$ to $\mathbf{z}_{r}$:
\vspace{-0.1cm}
\begin{equation} \mathbf{z}_{r} = \mathbf{z}_{t} + (r-t)\, \mathbf{u}_{\theta}(\mathbf{z}_{t}, t, r, \mathbf{E}) \label{eq:meanflow_update} \end{equation} 


\noindent MeanFlow-TSE~\cite{MeanFlowTSE} applies Eq.~\eqref{eq:meanflow_update} to the AD-FlowTSE \cite{ADFlowTSE} path with $t = \hat{\lambda}$ and $r = 1$, extracting the target in one step:
\vspace{-0.1cm}
\begin{equation} \hat{\mathbf{S}} = \mathbf{Y} + (1-\hat{\lambda})\, \mathbf{u}_{\theta}(\mathbf{Y}, \hat{\lambda}, 1, \mathbf{E}) \label{eq:meanflow_infer} \end{equation}

\noindent To train $\mathbf{u}_{\theta}$ efficiently, MeanFlow-TSE follows the $\alpha$-Flow framework~\cite{Alphaflow}, in which the network is trained toward a hybrid target $\mathbf{u}^{\alpha}_{t,r}$ that interpolates between the ground-truth direction and a self-consistency target. With the ground-truth direction $\mathbf{v}_{t} = \mathbf{S} - \mathbf{B}$, the hybrid target is defined as:

\vspace{-0.2cm}
\begin{equation}
\mathbf{u}^{\alpha}_{t,r} = \alpha\, \mathbf{v}_{t} + (1-\alpha)\, \mathrm{sg}\!\left[ \mathbf{u}_{\theta}(\mathbf{z}_{\tau}, \tau, r, \mathbf{E}) \right]
\label{eq:alpha_target}
\end{equation}

\noindent where $\alpha \in (0,1]$ is a ratio, $\tau = \alpha r + (1-\alpha)t$ is an intermediate time, and $\mathrm{sg}(\cdot)$ is the stop-gradient operator. Writing the error vector as $\boldsymbol{\Delta} = \mathbf{u}_{\theta}(\mathbf{z}_{t}, t, r, \mathbf{E}) - \mathbf{u}^{\alpha}_{t,r}$, the objective is minimized under an adaptive weight $w = \alpha / (\|\boldsymbol{\Delta}\|^{2} + c)$ with a small constant $c$:
\vspace{-0.2cm}

\begin{equation}
\mathcal{L}_{\alpha} = \mathbb{E}_{t, r, \mathbf{B}, \mathbf{S}} \!\left[ \mathrm{sg}(w)\, \big\| \boldsymbol{\Delta} \big\|^{2} \right]
\label{eq:lalpha}
\end{equation}

\noindent Annealing $\alpha$ from $1$ toward $0$ over training interpolates the objective from trajectory flow matching to the mean-flow identity. Despite its one-step inference, MeanFlow-TSE still suffers from slow training
convergence, requiring 2,000 epochs, which we aim to reduce. 

\vspace{-0.2cm}

 \begin{figure}[t]
  \centering
  \includegraphics[width=0.9\linewidth]{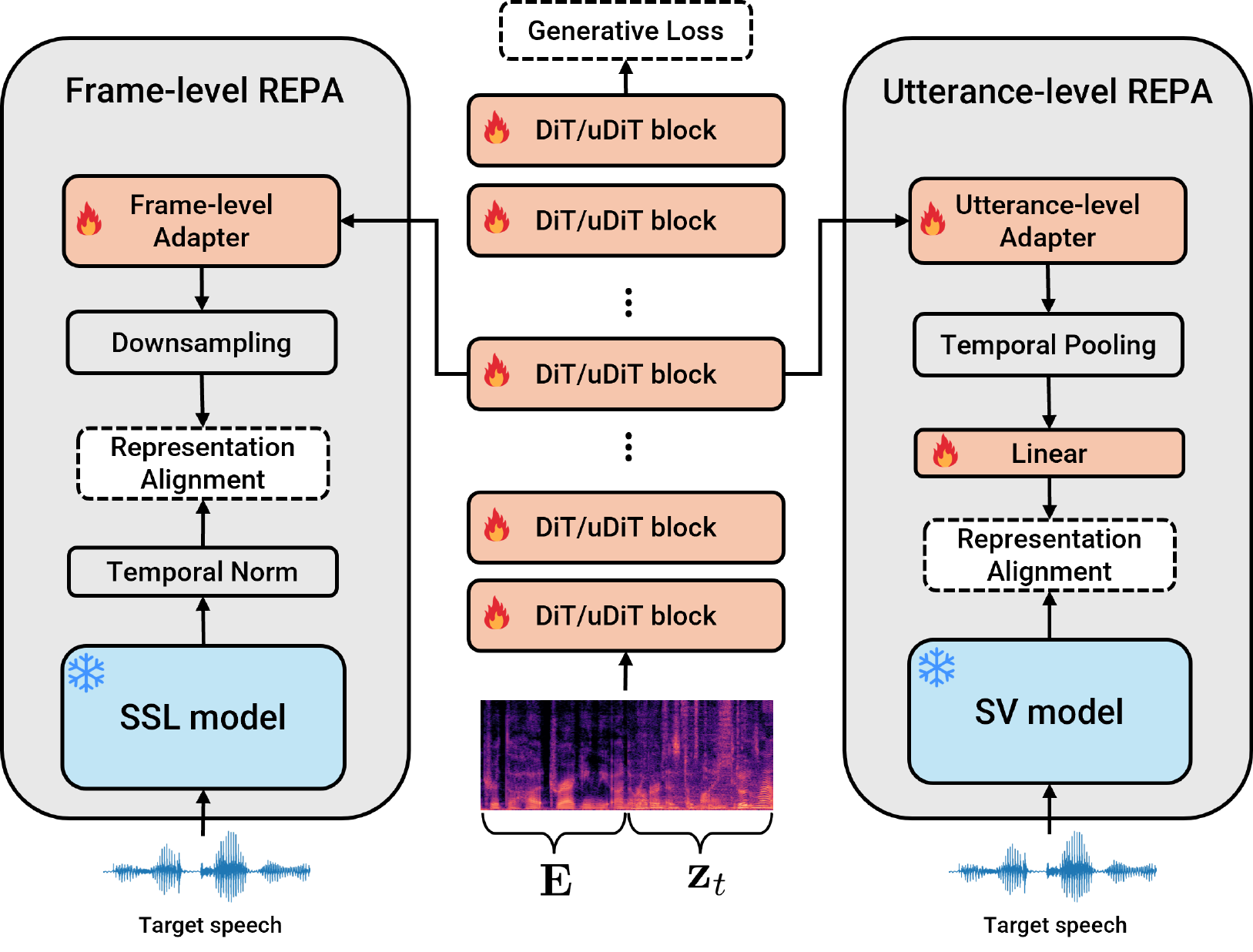}
  \caption{Overview of MG-REPA.}
  \label{fig:overview}
  \vspace{-0.5cm}
\end{figure}
\section{MG-REPA}
\label{sec:mgrepa}

MG-REPA supervises a single intermediate layer of the diffusion transformer at two granularities:
~\textbf{the frame-level alignment} for fine-grained phonetic content, and \textbf{the utterance-level alignment} for the global, time-invariant characteristics of the target speech, as illustrated in Fig.~\ref{fig:overview}.
MG-REPA extends REPA~\cite{REPA}, a method that accelerates diffusion transformer training by using a lightweight trainable adapter to match the hidden states from an intermediate block to the representations produced by a frozen pre-trained SSL teacher from the clean sample, and by optimizing this alignment loss jointly with the generative objective.

MG-REPA keeps this recipe but decouples the alignment into two branches, each with its own teacher and adapter. Since each DiT token corresponds to one STFT frame, let $\mathbf{H}^{(l)} \in \mathbb{R}^{T \times D}$ denote the hidden states of the $l$-th block over the mixture segment, computed from $\mathbf{z}_{t}$, with embedding dimension $D$. Both alignment branches take $\mathbf{H}^{(l)}$ as input and align it, under the cosine similarity $\mathrm{sim}(\cdot,\cdot)$, with representations that frozen pre-trained teachers extract from the target speech $\mathbf{s}$: the frame-level REPA with a speech SSL teacher (Sec.~\ref{ssec:frame_repa}), and the utterance-level REPA with a speaker verification teacher (Sec.~\ref{ssec:utterance_repa}).

\vspace{-0.2cm}
\subsection{Frame-level REPA}
\vspace{-0.1cm}
\label{ssec:frame_repa}

\noindent\textbf{Frame-level SSL teacher.} As the frame-level teacher, we use a frozen SSL encoder, WavLM~\cite{WAVLM}, whose upper layers are known to encode phonetic content~\cite{COMPARATIVE}. Let $\mathbf{f}^* = \mathcal{E}^{(k)}_{\mathrm{SSL}}(\mathbf{s}) \in \mathbb{R}^{T' \times D_{\mathrm{SSL}}}$ be the layer-$k$ features of WavLM for the target speech, where $T' < T$ because the encoder operates at a lower frame rate than the STFT.

\vspace{0.1cm}
\noindent\textbf{Frame-level adapter and temporal normalization.} Before alignment, the branch matches the hidden states $\mathbf{H}^{(l)}$ to the teacher frame rate and removes utterance-level statistics from the target. On the prediction side, an MLP adapter $h_\psi: \mathbb{R}^{D} \to \mathbb{R}^{D_{\mathrm{SSL}}}$ maps each frame of $\mathbf{H}^{(l)}$ to the teacher dimension, and $\mathrm{ds}: \mathbb{R}^{T \times D_{\mathrm{SSL}}} \to \mathbb{R}^{T' \times D_{\mathrm{SSL}}}$ downsamples the result to the teacher frame rate by adaptive average pooling along the time axis, yielding the prediction $\hat{\mathbf{f}} = \mathrm{ds}\big(h_\psi(\mathbf{H}^{(l)})\big)$. On the target side, $\mathbf{f}^*$, if left unnormalized, would retain its global statistics, which would make this branch overlap with the utterance-level branch (Sec.~\ref{ssec:utterance_repa}). 
Following~\cite{chou2019one}, which applies instance normalization to remove speaker information from content representations in voice conversion, we apply temporal normalization (TN) to the teacher features, normalizing each channel $d$ over the frames $n$:

\vspace{-0.4cm}
\begin{equation}
\mathrm{TN}\big(\mathbf{f}^*\big)_{n,d} =
\frac{f^*_{n,d} - \mu_{d}}{\sigma_{d} + \epsilon},
\qquad
\mu_{d} = \frac{1}{T'}\sum_{n=1}^{T'} f^*_{n,d},
\label{eq:temporal_norm}
\end{equation}
with $\sigma_{d}$ the corresponding standard deviation over the $T'$ frames and $\epsilon$ a small constant for stability. Removing the temporal mean and standard deviation of each channel suppresses the global, time-invariant information carried by these statistics while preserving the time-varying structure of the features. Since the normalized target carries little utterance-level information, the frame-level branch concentrates on local, fine-grained structure and leaves global characteristics to the utterance-level branch. This design is in line with prior work showing that removing speaker information from SSL features improves performance on content-oriented tasks~\cite{qian2022contentvec}.

\vspace{0.1cm}

\noindent\textbf{Frame-level alignment loss.} We define $\mathcal{L}_{\mathrm{frame}}$ as the negative frame-wise cosine similarity between the normalized target $\mathrm{TN}(\mathbf{f}^*)$ and the prediction $\hat{\mathbf{f}}$:

\begin{equation}
\mathcal{L}_{\mathrm{frame}} = -\, \mathbb{E} \! \bigg[ \frac{1}{T'} \sum_{n=1}^{T'} \mathrm{sim}\!\left( \mathrm{TN}\big(\mathbf{f}^*\big)_{n},\; \hat{\mathbf{f}}_{n} \right) \bigg].
\label{eq:frame_repa}
\end{equation}

\subsection{Utterance-level REPA}
\label{ssec:utterance_repa}

\noindent\textbf{Utterance-level SV teacher.} As the utterance-level teacher, we use a frozen speaker verification (SV) model, CAM++~\cite{CAMpp}, which aggregates its frame-level features into a single, time-invariant speaker embedding by statistics pooling. Let $\mathbf{g}^* = \mathcal{E}_{\mathrm{SV}}(\mathbf{s}) \in \mathbb{R}^{D_{\mathrm{SV}}}$ be the speaker embedding of the target speech. This embedding supervises the global speaker characteristics, including those removed from the frame-level target by the temporal normalization of Eq.~\eqref{eq:temporal_norm}.

\vspace{0.1cm}
\noindent\textbf{Utterance-level adapter and temporal pooling.} Since the target $\mathbf{g}^*$ is a single vector per utterance, the branch aggregates the $T$ hidden states into a single prediction through a gated adapter, temporal average pooling, and a linear projector. The adapter $h_\phi: \mathbb{R}^{T \times D} \to \mathbb{R}^{T \times D}$ is a gated linear unit~\cite{Gated} that multiplies a frame-wise MLP output by a sigmoid gate computed with a 1-D conv along the time axis. Its output $\tilde{\mathbf{H}} = h_\phi\big(\mathbf{H}^{(l)}\big)$ is averaged over time and mapped by the linear projector $p_\omega: \mathbb{R}^{D} \to \mathbb{R}^{D_{\mathrm{SV}}}$ to yield the prediction
\begin{equation}
\hat{\mathbf{g}} = p_\omega\Big(\frac{1}{T}\sum_{n=1}^{T}\tilde{\mathbf{H}}_n\Big).
\label{eq:pool}
\end{equation}

\noindent\textbf{Utterance-level alignment loss.} We define $\mathcal{L}_{\mathrm{utt}}$ as the negative cosine similarity between $\mathbf{g}^*$ and $\hat{\mathbf{g}}$:
\begin{equation}
\mathcal{L}_{\mathrm{utt}} = -\,\mathbb{E}\big[\mathrm{sim}(\mathbf{g}^*, \hat{\mathbf{g}})\big].
\label{eq:utt}
\end{equation}
Unlike $\mathcal{L}_{\mathrm{frame}}$, which averages the similarity over $T'$ frames, $\mathcal{L}_{\mathrm{utt}}$ compares a single pair of vectors per utterance.

\vspace{-0.1cm}
\subsection{Training Objective} \label{ssec:objective}
Given the generative objective $\mathcal{L}_{\mathrm{gen}}$ of the backbone, the generator and the alignment branches are trained jointly by minimizing
\begin{equation}
\mathcal{L} = \mathcal{L}_{\mathrm{gen}} + \beta_{\mathrm{f}}\, \mathcal{L}_{\mathrm{frame}} + \beta_{\mathrm{u}}\, \mathcal{L}_{\mathrm{utt}},
\label{eq:total_loss}
\end{equation}
where $\beta_{\mathrm{f}}$ and $\beta_{\mathrm{u}}$ weight the frame- and utterance-level alignment losses, while both teachers remain frozen. Since neither alignment loss depends on the prior, the path endpoints, or the regression target, $\mathcal{L}_{\mathrm{gen}}$ can be any generative loss, such as the flow matching loss of Eq.~\eqref{eq:fm} or the $\alpha$-Flow objective of Eq.~\eqref{eq:lalpha}. In this work, we apply MG-REPA to MeanFlow-TSE with $\mathcal{L}_{\mathrm{gen}} = \mathcal{L}_{\alpha}$. At inference, both branches and their teachers are discarded, so the inference cost is identical to that of the backbone.

\vspace{-0.2cm}
\section{Experiments}
\vspace{-0.2cm}
\subsection{Datasets}
\vspace{-0.1cm}
\label{ssec:datasets}

We follow the setup of~\cite{delcroix2020improving} and use Libri2Mix, a LibriSpeech-derived corpus built for multi-speaker separation and extraction, for both training and evaluation. The \texttt{train-100} and \texttt{train-360} subsets are combined for training, while \texttt{dev} and \texttt{test} are reserved for validation and evaluation, respectively. Inputs are 6-second segments at 16~kHz, formed by a 3-second enrollment and a 3-second mixture, and the spectral features are obtained with an STFT using a window length and FFT size of 510 and a hop length of 128.

\vspace{-0.15cm}
\subsection{Training Configuration}
\vspace{-0.1cm}

\label{ssec:impl}

\noindent\textbf{Backbone and training.} As noted in Sec.~\ref{ssec:objective}, MG-REPA constrains only an intermediate representation of the generator, so it is independent of the generative objective and can be applied to any backbone that exposes frame-indexed hidden states. We instantiate it on the UNet-style DiT (uDiT) velocity estimator shared by AD-FlowTSE and MeanFlow-TSE, which adopt the same architecture and differ only in their training objectives. We report results with $\mathcal{L}_{\mathrm{gen}}=\mathcal{L}_{\alpha}$ under the recipe of MeanFlow-TSE, following its official implementation. The estimator comprises 8 encoder blocks, a bottleneck block, and 8 decoder blocks linked to the encoder blocks by skip connections, each with 16 attention heads and a hidden size of 1024. We optimize with AdamW~\cite{ADAMW} at a weight decay of 0.01, an initial learning rate of $1\times10^{-4}$ annealed by a cosine policy with $T_{\max}=50$, a minimum learning rate of $1\times10^{-5}$, and a 5-epoch warm-up. Gradients are clipped at 0.5, and 16-bit mixed-precision training is used under distributed data parallelism with a batch size of 256.

Both MG-REPA branches are attached to the output of the bottleneck block ($l\!=\!9$). The frame-level branch uses the 24th layer ($k\!=\!24$) of frozen WavLM Large, and the utterance-level branch uses the output of frozen CAM++. The loss weights in Eq.~\eqref{eq:total_loss} are set to $\beta_{\mathrm{f}} = \beta_{\mathrm{u}} = 0.5$.


\begin{table*}[t]
\centering
\begin{threeparttable}
\fontsize{7.5pt}{10pt}\selectfont
\caption{Performance comparison of generative TSE models on the Libri2Mix Clean and Noisy sets.}
\label{tab:results}
\begin{tabular}{l cccccc cccccc}
\toprule
\multirow{2}{*}{Method} & \multicolumn{6}{c}{Libri2Mix Clean} & \multicolumn{6}{c}{Libri2Mix Noisy} \\
\cmidrule(lr){2-7} \cmidrule(lr){8-13}
 & PESQ & ESTOI & SI-SDR & OVRL & DNSMOS & SIM & PESQ & ESTOI & SI-SDR & OVRL & DNSMOS & SIM \\
\midrule
DiffSep+SV~\cite{DDTSE} & 1.85 & 0.79 & -- & 3.14 & \textbf{3.83} & 0.83 & 1.32 & 0.60 & -- & 2.78 & 3.63 & 0.62 \\
DDTSE~\cite{DDTSE} & 1.79 & 0.78 & -- & \textbf{3.30} & 3.79 & 0.73 & 1.60 & 0.71 & -- & 3.28 & 3.74 & 0.71 \\
DiffTSE~\cite{DiffTSE2} & 3.08 & 0.80 & 11.28 & -- & -- & -- & -- & -- & -- & -- & -- & -- \\
FlowTSE~\cite{flowtse} & 2.58 & 0.84 & -- & 3.27 & 3.79 & 0.90 & 1.86 & 0.75 & -- & \textbf{3.30} & \textbf{3.82} & 0.83 \\
SR-SSL~\cite{SRSSL} & 2.99 & -- & 16.00 & -- & -- & -- & -- & -- & -- & -- & -- & -- \\
SoloSpeech~\cite{SOLOSPEECH} & -- & -- & -- & -- & -- & -- & 1.89 & 0.78 & 11.12 & -- & 3.76 & -- \\
AD-FlowTSE~\cite{ADFlowTSE} & 2.89 & 0.90 & 17.49 & 3.15 & 3.59 & \textbf{0.95} & 2.15 & 0.81 & 12.69 & 3.11 & 3.48 & \textbf{0.87} \\
\midrule
MeanFlow-TSE~\cite{MeanFlowTSE} {\fontsize{6}{6}\selectfont (2,000 epochs)} & 3.26 & \textbf{0.93} & 18.80 & 3.21 & 3.69 & 0.92 & 2.21 & \textbf{0.82} & \textbf{12.85} & 3.17 & 3.55 & 0.73 \\
\qquad + MG-REPA {\fontsize{6}{6}\selectfont ~~~~~~(750 epochs)} & \textbf{3.29} & \textbf{0.93} & \textbf{18.88} & 3.24 & 3.73 & 0.92 & \textbf{2.22} & \textbf{0.82} & 12.38 & 3.20 & 3.60 & 0.73 \\
\bottomrule
\end{tabular}

\vspace{-0.3cm}
\end{threeparttable}
\end{table*}

\noindent\textbf{Baseline Methods.} We compare MG-REPA with generative TSE baselines: DiffSep+SV and DDTSE~\cite{DDTSE}, DiffTSE~\cite{DiffTSE2}, FlowTSE~\cite{flowtse}, SR-SSL~\cite{SRSSL}, SoloSpeech~\cite{SOLOSPEECH}, AD-FlowTSE~\cite{ADFlowTSE}, and MeanFlow-TSE~\cite{MeanFlowTSE}, which serves as our backbone. 

\vspace{0.1cm}
\noindent\textbf{Evaluation Metrics.}
For extensive validation, we evaluate signal fidelity and intelligibility with SI-SDR~\cite{SISDR}, PESQ~\cite{PESQ}, and ESTOI~\cite{ESTOI}. Non-intrusive quality is measured with DNSMOS~\cite{DNSMOS} and its OVRL score. Speaker similarity (SIM) is the cosine similarity between speaker embeddings~\cite{WESPEAKER} of the estimated and target speech. For the ablation on alignment granularity in Sec.~\ref{ssec:ablation}, we additionally report Levenshtein phoneme similarity (LPS)~\cite{LPS} to verify that phonetic content is preserved, and UTMOS~\cite{UTMOS} as a predicted naturalness score. Higher is better for all metrics.

\vspace{-0.2cm}
\section{Results}
\vspace{-0.1cm}

\subsection{Overall Performance}
\vspace{-0.1cm}

\label{ssec:overall}
Table~\ref{tab:results} compares MG-REPA with generative TSE baselines on the clean and noisy subsets of Libri2Mix. MG-REPA is trained for 750 epochs, whereas the MeanFlow-TSE backbone is trained for 2,000 epochs. Despite the reduced budget, MG-REPA matches or exceeds the backbone on every metric except noisy SI-SDR, reaching its quality with $2.7\times$ fewer training epochs.

On the clean set, MG-REPA achieves the best PESQ (3.29) and SI-SDR (18.88~dB) among the methods that report them, and already attains the 2,000-epoch ESTOI of the backbone (0.93). On the noisy set, it achieves the best PESQ (2.22) and again attains the backbone ESTOI (0.82). The only metric that falls below the backbone is noisy SI-SDR ($-$0.47~dB). Since both teachers extract their targets from the clean target speech, the alignment has to bridge a larger gap between the mixture hidden states and the teacher representations under additive noise, and noisy SI-SDR is the only metric for which 750 epochs do not recover the 2,000-epoch level.

MG-REPA improves non-intrusive scores on both sets: on clean, OVRL +0.03 and DNSMOS +0.04; on noisy, OVRL +0.03 and DNSMOS +0.05. Speaker similarity also reaches the 2,000-epoch level of the backbone. SIM is computed with WeSpeaker embeddings~\cite{WESPEAKER}, not the CAM++ teacher in Sec.~\ref{ssec:utterance_repa}, so the utterance-level branch does not directly optimize SIM; its benefit appears in non-intrusive scores, consistent with the ablation of Sec.~\ref{ssec:ablation}. DiffSep+SV, DDTSE, and FlowTSE still lead on non-intrusive metrics, reflecting the intrusive–non-intrusive trade-off reported in~\cite{ADFlowTSE, MeanFlowTSE}, but MG-REPA narrows the gap.

\vspace{-0.4cm}
\subsection{Model Convergence Speed} \label{ssec:convergence}
\vspace{-0.1cm}
Fig.~\ref{fig:convergence} compares the OVRL and PESQ of MG-REPA and MeanFlow-TSE on the Libri2Mix Noisy set over 750 training epochs. Since the MeanFlow-TSE results in Table~\ref{tab:results} are taken from the original paper, we retrain MeanFlow-TSE for this comparison with the recipe of Sec.~\ref{ssec:impl} under the same 750-epoch budget as MG-REPA. Fig.~\ref{fig:convergence} shows that MG-REPA reaches the 750-epoch level of this retrained baseline at epoch 250 for OVRL and 350 for PESQ, i.e., with $3.0\times$ and $2.1\times$ fewer training epochs, and the gap to the baseline persists throughout training.
\vspace{-0.15cm}

\vspace{-0.2cm}
\subsection{Ablation on Alignment Granularity}
\vspace{-0.1cm}
\label{ssec:ablation}

Table~\ref{tab:ablation_granularity} isolates the contribution of each alignment granularity. Aligning the frame level alone with WavLM (b) already improves every metric over the MeanFlow-TSE baseline (a), and replacing this branch with utterance-level supervision from CAM++ (c) yields a comparable gain, so each granularity carries supervision that the baseline lacks.

\noindent We then use both teachers, keeping the teachers, their layers, and the training scheme fixed so that (d)--(f) differ only in granularity. Aligning both at the frame level (d), with CAM++ features taken before statistics pooling as time-varying cues in TSE~\cite{MultiSV}, improves over frame-level WavLM alone (b) on every metric except ESTOI.
Aligning both at the utterance level (e) averages the WavLM features over time, following the temporal pooling of~\cite{AG-REPA}, and reduces them to their temporal mean $\mu$. This raises ESTOI over CAM++ alone (c), leaves LPS unchanged, and lowers PESQ, SI-SDR, OVRL, and UTMOS. Assigning each teacher to its own granularity (f) matches or exceeds both single-granularity combinations (d) and (e) on every metric. Since (e) and (f) differ only in the granularity of the WavLM target, the benefit of this target comes mainly from its time-varying structure rather than from $\mu$, the utterance-level mean.





TN removes the utterance-level statistics from the frame-level target so that each branch focuses on its own granularity.
Applying TN to the WavLM target (g) leaves ESTOI and LPS unchanged and improves every other metric over (f), achieving the best performance.
Compared with aligning both teachers at the frame level (d), (g) improves PESQ by 0.02, SI-SDR by 0.09~dB, ESTOI by 0.01, OVRL by 0.03, and UTMOS by 0.04. The gain from multi-granularity alignment is not a trade-off between signal fidelity and perceived naturalness, but a consistent improvement in both.

\begin{table}[t]

\centering
\fontsize{8pt}{10pt}\selectfont
\setlength{\tabcolsep}{2.6pt}
\caption{Ablation on alignment granularity on Libri2Mix Noisy. 
All variants are trained for 150 epochs. W: WavLM, C: CAM++.}




\label{tab:ablation_granularity}
\begin{tabular}{@{}c ccc cccccc@{}}
\toprule
& \multicolumn{3}{c}{Alignment} & & & & & & \\
\cmidrule(lr){2-4}
& Frame & Utt. & TN & PESQ & ESTOI & SI-SDR & OVRL & UTMOS & LPS \\
\midrule
(a) & --      & --      & --         & 1.74 & 0.74 & 10.34 & 2.93 & 2.88 & 0.80 \\
\midrule
(b) & W       & --      & --         & 1.94 & 0.77 & 11.35 & 3.06 & 3.25 & 0.85 \\
(c) & --      & C       & --         & 1.92 & 0.76 & 11.30 & 3.07 & 3.23 & 0.84 \\
\midrule
(d) & W\,+\,C & --      & --         & 1.97 & 0.77 & 11.51 & 3.07 & 3.28 & \textbf{0.86} \\
(e) & --      & W\,+\,C & --         & 1.91 & \textbf{0.78} & 11.09 & 3.05 & 3.22 & 0.84 \\
\midrule
(f) & W       & C       & --         & 1.97 & \textbf{0.78} & 11.55 & 3.09 & 3.29 & \textbf{0.86} \\
\rowcolor{gray!12}
(g) & W       & C       & \checkmark & \textbf{1.99} & \textbf{0.78} & \textbf{11.60} & \textbf{3.10} & \textbf{3.32} & \textbf{0.86} \\
\bottomrule
\vspace{-0.8cm}

\end{tabular}
\end{table}

\vspace{-0.3cm}
\section{Conclusion} \label{sec:conclusion} 
\vspace{-0.2cm}

We proposed MG-REPA, which aligns an intermediate layer of a generative TSE model with temporally normalized WavLM features at the frame level and a CAM++ speaker embedding at the utterance level. On Libri2Mix, MG-REPA reaches the OVRL and PESQ of the MeanFlow-TSE backbone with $3.0\times$ and $2.1\times$ fewer epochs, respectively. Ablations show that supervising both granularities outperforms alignment at either granularity alone. Future work will extend MG-REPA to other generative TSE backbones, such as AD-FlowTSE, to verify its generality.

\section{Acknowledgement}
\vspace{-0.3cm}
This work was supported by the National Research Foundation of Korea (NRF) grant (No. RS-2024-00337945), the STEAM research grant (No. RS-2024-00464269) funded by the Ministry of Science and ICT of Korea government (MSIT), the BK21 FOUR program through the NRF grant funded by the Ministry of Education of Korea government (MOE), Artificial intelligence industrial convergence cluster development project funded by the Ministry of Science and ICT(MSIT, Korea) \& Gwangju Metropolitan City. Hyundai Motor Company and Kia (No. G01250273).

\vspace{-0.2cm}

\bibliographystyle{IEEEbib}
{\ninept
\bibliography{strings,refs}
}
\end{document}